\documentclass[10pt]{wlscirep}

\def \B{B_{\rm p}}

\usepackage{float}
\floatstyle{boxed}
\usepackage{wrapfig}
\usepackage{array}
\usepackage{graphicx}
\usepackage{amsbsy}
\usepackage{xcolor}
\usepackage{epstopdf}
\begin{document}

% \title{Uncovering nonclassicality of noisy twin beams on a unbalanced beam splitter with  photon-number statistics}
\title{Experimental identification of non-classicality of noisy
twin beams and other related two-mode states}
\author[1,*]{Ievgen I. Arkhipov}
\author[1]{Jan Pe\v{r}ina Jr.}
\affil{RCPTM, Joint Laboratory of Optics of Palack\'y University and
Institute of Physics of CAS, Faculty of Science, Palack\'y University, 17. listopadu
12, 771 46 Olomouc, Czech Republic}
\affil[*]{ievgen.arkhipov@gmail.com}

\begin{abstract}
Different non-classicality criteria expressed in the form of
inequalities among intensity moments and elements of photon-number
distributions are applied to noisy twin beams and other two-mode
states obtained from a twin beam by using a beam splitter. Their
performance in revealing the non-classicality is judged in
comparison with the exact results provided by suitable
entanglement and local non-classicality quantifiers. Whereas the
non-classicality of noisy twin beams is always revealed by these
criteria, not all the nonclassical states obtained at the output
of the beam splitter can be identified by these experimentally
easily reachable criteria.
\end{abstract}

\maketitle

\section*{Introduction}

Nonclassical properties of light have been in the focus of
investigations in quantum nonlinear optics for a long time. Broad
and deep studies of non-classicality of optical fields and
especially of entanglement among their parts even resulted in
establishing a new field in science --- quantum information
science~\cite{NielsenBook,Wilde2013}. Historically, discrete- and
continuous-variable quantum-optical systems have been
distinguished when investigating their nonclassical properties.
Recently, the so-called hybrid quantum-optical systems composed of
both discrete- and continuous-variable parts in the mutual
interaction have been addressed. Here, we pay attention to
non-classicality of a two-mode optical field that has probably
been the most frequently studied system described by the
continuous variables. It is well known that these fields can
exhibit both entanglement between their
modes~\cite{Braunstein05_RMP,Weedbrook12} and squeezing inside
their modes~\cite{MandelBook,Perina1991Book}. Whereas the
squeezing of the modes manifests local non-classicality of these
modes, the entanglement between the modes is responsible for
global non-classicality of the overall two-mode field. Moreover,
the global non-classicality of the overall field is also implied
by the local non-classicalities of individual modes~\cite{Kim2002}.

Different techniques have been developed to experimentally verify
non-classicality of optical fields. The most elaborated, and also
the most experimentally demanding, technique is homodyne
tomography that relies on homodyne detection \cite{Lvovsky2009}
when reconstructing a quantum
state~\cite{Shchukin2006,Sperling2012a}. On the other hand, the
usual quadratic optical detectors recording just the field's
intensity or the fields' intensity correlations have undergone
fast development in the last ten years and their recent variants
aimed at detecting weak optical fields provide photon-number
resolution. This brings the opportunity to identify
non-classicality under suitable conditions, even not knowing the
phases of fields' complex amplitudes. In principle, an
experimental photocount histogram can be used to reconstruct the
quasi-distribution of integrated intensities (or more exactly its
regularized form) and reveal the non-classicality through its
negative values \cite{Haderka2005a,PerinaJr2013a}. Or one can just
rely on the application of various non-classicality criteria
usually in the form of inequalities, as it has been done, e.g., in
Refs.~\cite{Arkhipov2016e,PerinaJr2017} for single-mode fields and
in Ref.~\cite{PerinaJr2017a} for two-mode fields.

The importance of identification of non-classicality through the
measurement with photon-number-resolving detectors is constantly
growing as these detectors are becoming more efficient and so more
popular. Superconducting bolometers are at present the most
sophisticated detectors of this kind. They are endowed with the
best quantum detection efficiencies, but at the expense of their
cryogenic operation. On the other hand, the oldest fiber-based
photon-number-resolving detectors with time multiplexing are
relatively cheap and easy to
operate~\cite{Achilles2003,Fitch2003,Haderka2004,Avenhaus2010,Sperling2015}.
Ultra-sensitive cameras \cite{Mosset2004,Blanchet2008}, especially
intensified CCD (iCCD)
cameras~\cite{Haderka2005a,PerinaJr2012,PerinaJr2013a,Arkhipov2016e,PerinaJr2017b},
can also be applied as massively spatially-multiplexed
photon-number-resolving detectors. Due to large numbers of pixels
on their photocathodes, these cameras are suitable also for
characterizing mesoscopic optical fields \cite{Machulka2014}.
Hybrid detectors \cite{PerinaJr2014a} and also silicon multi-pixel
detection arrays \cite{Ramilli2010} also belong to prospective
photon-number-resolving detectors at present.

When applying various non-classicality criteria an important
question about their performance arises. Namely, how many
nonclassical states can be revealed by these criteria. Also mutual
comparison of different criteria has important consequences for
their practical use. Many non-classicality inequalities applied to
specific kinds of both single mode and two-mode optical fields
have been mutually compared in
Refs.~\cite{Arkhipov2016e,PerinaJr2017,PerinaJr2017a} using the
experimental data. The applied non-classicality inequalities have
shown good experimental performance in the cases of sub-Poissonian
states and practically noiseless twin beams. This poses the
question about their power in the cases of other nonclassical
states that can be relatively easily reached in the laboratory.
Here, we theoretically address two such kinds of states: noisy
twin beams and two-mode states derived from the noisy twin beams
using, in general, an un-balanced beam splitter. Especially the
second kind of states is interesting as it allows, via the
bunching effect of photon pairs at a beam splitter, to obtain
squeezed single-mode
states~\cite{Paris97,Kim2002,Braunstein05_RMP,Weedbrook12}, that
exhibit local non-classicality. To simulate the experimental
identification of non-classicality we apply the non-classicality
criteria written in either the normally-ordered intensity moments
or the elements of photon-number (photocount) distributions that
have been developed and summarized in Ref~\cite{PerinaJr2017a}. To
theoretically judge local non-classicality and entanglement of the
analyzed states we invoke the local non-classicality and
entanglement quantifiers derived in
Refs.~\cite{Arkhipov2016c,Arkhipov2016b}, that are especially
suitable for two-mode Gaussian states and their transformations at
a beam splitter. We show that whereas the applied global
non-classicality criteria (GNCCa) allow us to recognize all
entangled noisy twin beams, not all two-mode nonclassical states
occurring beyond the beam splitter can be identified with the used
GNCCa and local non-classicality criteria (LNCCa). Our analysis
also shows that the criteria based on the elements of
photon-number distributions exhibit better performance compared to
those written in intensity moments. We also identify the most
powerful criteria.

The paper is organized as follows. Section {\it Two-mode optical
fields and their properties} brings the description of
nonclassical properties of optical fields. The analyzed
non-classicality criteria are mentioned in section {\it
Non-classicality criteria}. The states originating in the noisy
twin beams propagating through an un-balanced beam splitter are
discussed in section {\it Twin beam and its transformation on a
beam splitter}. Section {\it Identification of non-classicality of
twin beams} is devoted to the performance of the used
non-classicality criteria in revealing the entanglement of noisy
twin beams. Similarly, different non-classicality criteria are
applied in section {\it Identification of non-classicality of
two-mode states beyond a beam splitter} to reveal non-classicality
of the addressed states. In section {\it Conclusions} the main
findings are summarized.

\section*{Two-mode optical fields and their properties}

Any two-mode state characterized by its density matrix $\hat\rho$
can be expressed in the Glauber-Sudarshan (diagonal)
representation based on coherent states $|\alpha_1\rangle$ and
$|\alpha_2\rangle$ defined in modes 1 and 2, respectively:
\begin{equation}  % 1
 \hat\rho=\frac{1}{\pi^2}\int d^2\alpha_1 \int d^2\alpha_2
  {\cal P}(\alpha_1,\alpha_2)|\alpha_1\rangle|\alpha_2\rangle\langle\alpha_2|\langle\alpha_2|.
\label{1}
\end{equation}
In Eq.~(\ref{1}), $ {\cal P}(\alpha_1,\alpha_2)$ stands for the
Glauber-Sudarshan
quasi-distribution~\cite{Glauber1963,Sudarshan1963}. The
quasi-distribution $ {\cal P} $ uniquely identifies nonclassical
states. If it attains the form of a regular distribution function
with non-negative values (or has the form of a sum of the Dirac $
\delta $-functions) it describes a classical state. However, if it
becomes negative or even more singular than the Dirac $ \delta
$-function, it corresponds to a nonclassical state.

If the information about the phase of an optical field is not
known, we can restrict our attention to quasi-distribution $
P(W_1,W_2) $ of integrated intensities $ W_1 $ and $ W_2 $ ($ W_j
= |\alpha_j|^2 $, $ j=1,2 $) \cite{Perina1991Book}, instead of using
the Glauber-Sudarshan quasi-distribution $ {\cal
P}(\alpha_1,\alpha_2) $. Moments $ \langle W_1^{k_1} W_2^{k_2}
\rangle $, $ k_1,k_2 =0,1,\ldots $, of the integrated intensities
$ W_1 $ and $ W_2 $ (farther only intensities) are then easily
determined by averaging with the intensity quasi-distribution $
P(W_1,W_2) $:
\begin{equation} % 2
 \langle W_1^{k_1} W_2^{k_2} \rangle = \int_{0}^{\infty} dW_1
 \int_{0}^{\infty} dW_2\, P(W_1,W_2)  W_1^{k_1} W_2^{k_2}.
\label{2}
\end{equation}
We note that the intensity moments $ \langle W^k \rangle $ are
just the normally ordered moments $ \hat{a}^{\dagger k} \hat{a}^k
$ of the photon-number operator as $ \langle W^k \rangle \equiv
\langle \hat{a}^{\dagger k} \hat{a}^k \rangle $ and $
\hat{a}^{\dagger} $ ($ \hat{a} $) stands for the creation
(annihilation) operator.

According to the Mandel photodetection formula~\cite{Perina1991Book},
photon-number distribution $ p(n_1,n_2) $ for a field with
intensity quasi-distribution $ P(W_1,W_2) $ is obtained as
follows:
\begin{equation} % 3
 p(n_1,n_2) = \frac{1}{n_1!n_2!} \int_{0}^{\infty} dW_1
  \int_{0}^{\infty} dW_2 \, P(W_1,W_2) W_1^{n_1}  W_2^{n_2} \exp[ -(W_1 + W_2)] .
\label{3}
\end{equation}

Both the intensity moments given in Eq.~(\ref{2}) and the elements
of photon-number distribution $ p $ written in Eq.~(\ref{3}) can
conveniently be derived from the normal generating function $
G_{\cal N} $ defined as:
\begin{equation} % 4
 G_{\cal N}(\lambda_1,\lambda_2) = \int_{0}^{\infty} dW_1
  \int_{0}^{\infty} dW_2 \, P(W_1,W_2)\exp( -\lambda_1 W_1 - \lambda_2 W_2) .
\label{4}
\end{equation}
Whereas the intensity moments $ \langle W_1^{k_1} W_2^{k_2}\rangle
$ are obtained along the formula
\begin{eqnarray} % 5
 \langle W_1^{k_1} W_2^{k_2}\rangle = (-1)^{k_1+k_2} \left.\frac{\partial^{k_1+k_2}
  G_{\cal
  N}(\lambda_1,\lambda_2)}{\partial\lambda_1^{k_1}\partial\lambda_2^{k_2}}\right|_{\lambda_1=\lambda_2=0},
\label{5}
\end{eqnarray}
the elements $ p(n_1,n_2) $ of photon-number distribution are
reached as follows:
\begin{eqnarray} % 6
 p(n_1,n_2) = \frac{ (-1)^{n_1+n_2} }{n_1!n_2!}
 \left.\frac{\partial^{n_1+n_2} G_{\cal
  N}(\lambda_1,\lambda_2)}{\partial\lambda_1^{n_1}\partial\lambda_2^{n_2}}\right|_{\lambda_1=\lambda_2=1}.
\label{6}
\end{eqnarray}

\section*{Non-classicality criteria}

We describe non-classicality criteria that have been derived in
Ref.~\cite{PerinaJr2017a} and have shown the best performance. The
violation of the classical inequality
\begin{equation}  % 7
 \langle W_1^{k_1} W_2^{k_2}(W_1-W_2)^2\rangle < 0
\end{equation}
gives us the following GNCCa for $ k_1, k_2 \ge 0 $:
\begin{equation} % 8
 E_{k_1,k_2}^W\equiv \langle W_1^{k_1+2} W_2^{k_2}\rangle + \langle W_1^{k_1} W_2^{k_2+2} \rangle
  -2\langle W_1^{k_1+1} W_2^{k_2+1} \rangle < 0.
\label{8}
\end{equation}
Following the correspondence between the GNCCa based on intensity
moments and the GNCCa containing the elements of photon-number
distribution discussed in Ref.~\cite{PerinaJr2017a} we arrive at the
following GNCCa:
\begin{equation} % 9
 E_{k_1,k_2}^p\equiv \tilde p(k_1+2,k_2) + \tilde p(k_1,k_2+2)
  -2\tilde p(k_1+1,k_2+1) < 0
\label{9}
\end{equation}
using the modified elements $ \tilde p(n_1,n_2) $ of photon-number
distribution,
\begin{equation}  % 10
 \tilde p(n_1,n_2) = \frac{ n_1! n_2! \,p(n_1,n_2) }{p(0,0)} .
\label{10}
\end{equation}

Also the GNCCa $ M^W $ and $ M^p $ defined along the relations
\begin{eqnarray} % 11
 & M^W \equiv \langle W_1^{2}\rangle \langle W_2^2\rangle - \langle W_1W_2\rangle^2 < 0, & \nonumber \\
 & M^p \equiv \tilde p(2,0)\tilde p(0,2) - \tilde p(1,1)^2 < 0 &
\label{11}
\end{eqnarray}
have been found powerful in Ref.~\cite{PerinaJr2017a} when revealing
non-classicality. We note that they originate in the matrix
approach \cite{Shchukin2005,Miranowicz2010,Bartkowiak2011b} that is based upon
non-negativity of classical quadratic forms.

The most powerful single-mode LNCCa have been derived in
Ref.~\cite{Lee1990a} using the majorization theory. They have been
tested on the experimental sub-Poissonian fields in
Ref.~\cite{Arkhipov2016e}. They attain the following form for mode
$ j $, $ j=1,2 $:
\begin{eqnarray} % 12
 & R_{k,l}^{W_j} \equiv \langle W_j^{k+1} \rangle \langle W_j^{l-1}\rangle - \langle W_j^{k}\rangle \langle W_j^{l} \rangle < 0,& \nonumber \\
 & R_{k,l}^{p_j} \equiv \tilde p_j(k+1)\tilde p_j(l-1)-\tilde p_j(k)\tilde p_j(l) < 0. &
\label{12}
\end{eqnarray}
The modified elements $ \tilde p_j(n) $ of marginal photon-number
distribution $ p_j(n) $ of mode $ j $ are given as $ \tilde p_j(n)
= n! p_j(n)/p_j(0) $.

\section*{Twin beam and its transformation on a beam splitter}

A twin beam is generated in the process of spontaneous parametric
down-conversion that generates photon pairs at the expense of
annihilated pump photons \cite{MandelBook}. Twin beams in general
contain more photon pairs and they can also contain an additional
noise in the form of individual photons~\cite{Perina1991Book}. Such
general noisy twin beams belong to two-mode Gaussian optical
fields that can be conveniently described by the normal quantum
characteristic function $ C_{\cal N} $ defined
as~\cite{Perina1991Book}
\begin{eqnarray}    % 13
 C_{\cal N}(\beta_{1},\beta_{2})=\left\langle\exp (\beta_{1}
  \hat a^{\dagger}_{1}+\beta_{2}\hat a^{\dagger}_{2})
  \exp(-\beta^{\ast}_{1}\hat a_{1} -\beta^{\ast}_{2}\hat
  a_{2})\right\rangle  \nonumber \\
= \int d^2\alpha_1 \int d^2\alpha_2
  \prod_{j=1}^2 \exp(\beta_j\alpha_j^* - \beta_j^*\alpha_j)
  {\cal P}(\alpha_1,\alpha_2)\hspace{-10mm}
\label{13}
\end{eqnarray}
using the Glauber-Sudarshan quasi-distribution $ {\cal P} $.

Both the noisy twin beams and the states arising beyond a beam
splitter with an impinging twin beam belong to two-mode Gaussian
states with the following form of the normal characteristic
function $ C_{\cal N} $:
\begin{equation}   % 14
 C_{\cal N}(\beta_1,\beta_2)= \exp\left[- B_1 |\beta_1|^2
  -B_2|\beta_2|^2+\left(\frac{C_1}{2}\beta^{*2}_1  +\frac{C_2}{2}\beta^{*2}_{2}+
 D_{12}\beta_1^*\beta_2^* + \bar{D}_{12}\beta_1\beta_2^*
 + {\rm c.c.} \right)\right];
\label{14}
\end{equation}
symbol c.c. replaces the complex-conjugated terms. The
coefficients $ B_j $, $ C_j $, $ j=1,2 $, $ D_{12} $, and $
\bar{D}_{12} $ introduced in Eq.~(\ref{14}) are defined as
follows:
\begin{eqnarray}  % 15
 B_j=\langle\Delta\hat a^{\dagger}_{j}\Delta\hat a_{j}\rangle, &\quad&  C_j=\langle(\Delta\hat a_{j})^2\rangle, \nonumber \\
 D_{12} =\langle\Delta\hat a_{1}\Delta\hat a_{2}\rangle, &\quad&  \bar{D}_{12} =-\langle\Delta\hat a_{1}^{\dagger}\Delta\hat
 a_{2}\rangle
\label{15}
\end{eqnarray}
and $ \Delta \hat a_j \equiv \hat a_j - \langle \hat a_j \rangle
$.

The normal characteristic function $ C_{\cal N} $ given in
Eq.~(\ref{14}) can conveniently be rewritten into the form $
C_{\cal N}(\mbox{\boldmath$ \beta $})= \exp (\mbox{\boldmath$
\beta $}^{\dagger}{\bf A}_{\cal N} \mbox{\boldmath$ \beta $}/2)$
using the covariance matrix $ {\bf A}_{\cal N} $ related to normal
ordering of the field
operators~\cite{PerinaKrepelka11,Perina1991Book},
\begin{eqnarray}   % 16
 {\bf A}_{\cal N} = \left[\begin{array}{cccc}
  - B_{1} & C_1 & {\bar{D}}_{12}^*& D_{12} \\
  C_{1}^* & -B_1 & D_{12}^* & {\bar{D}}_{12} \\
  {\bar{D}}_{12} &D_{12} &-B_2& C_2 \\
   D_{12}^* &  {\bar{D}}_{12}^* & C_{2}^* & -B_2
  \end{array}\right] ,
\label{16}
\end{eqnarray}
and the column vector $ \mbox{\boldmath$ \beta $} $ is given as $
\mbox{\boldmath$ \beta $} \equiv
(\beta_1,\beta_1^*,\beta_2,\beta_2^*)^T $.

The normally-ordered generating function $ G_{\cal N} $ from
Eq.~(\ref{4}) is then obtained along the
formula~\cite{PerinaKrepelka05}:
\begin{equation} % 17
 G_{\cal N}(\lambda_1,\lambda_2) =
  \frac{1}{\pi^2\lambda_1\lambda_2} \int d^2\beta_1 \int d^2\beta_2
  \exp\left(-\frac{|\beta_1|^2}{\lambda_1}-
  \frac{|\beta_2|^2}{\lambda_2}\right) C_{\cal N}(\beta_1,\beta_2).
\label{17}
\end{equation}
Considering the form of the characteristic function $ C_{\cal N} $
written in Eq.~(\ref{14}), we arrive at the generating function $
G_{\cal N} $ for the considered states:
\begin{eqnarray} % 18
 G_{\cal N}(\lambda_1,\lambda_2)= \frac{1}{ {[\mbox{\boldmath$ \lambda^{\rm
  T} $} {\bf K} \mbox{\boldmath$ \lambda $} ]^{1/2}} }
\label{18}
\end{eqnarray}
and $ \mbox{\boldmath$ \lambda $} \equiv
(1,\lambda_1,\lambda_2,\lambda_1\lambda_2)^{\rm T}$. The matrix $
{\bf K} $ occurring in Eq.~(\ref{18}) is obtained as
\begin{eqnarray}  % 19
&&\hspace{28mm}{\bf K} = \left[ \begin {array}{cccc} 1&{ K_{12}}&{ K_{13}}&{K_{14}}
  \\ \noalign{\medskip}0&{ K_{22}}&0&{ K_{24}}\\ \noalign{\medskip}0&0&{
   K_{33}}&{ K_{34}}\\ \noalign{\medskip}0&0&0&{ K_{44}}\end {array}
   \right],
\label{19}   \\
 & & K_{12}=2B_1, \nonumber \\
 & & K_{13}=2B_2, \nonumber \\
 & & K_{14}=4B_1B_2-2|D_{12}|^2-2|\bar D_{12}|^2, \nonumber \\
 & & K_{22}=B_1^2-|C_1|^2, \nonumber \\
 & & K_{24}=2B_1^2B_2-2B_1\left(|D_{12}|^2+|\bar D_{12}|^2\right) -2B_2|C_1|^2-4{\rm Re}\left[C_1\bar D_{12}D_{12}^{*}\right], \nonumber \\
 & & K_{33}=B_2^2-|C_2|^2, \nonumber \\
 & & K_{34}=2B_1B_2^2-2B_2\left(|D_{12}|^2+|\bar D_{12}|^2\right) -2B_1|C_2|^2-4{\rm Re}\left[C_2\bar D_{12}^{*}D_{12}^{*}\right], \nonumber \\
 & & K_{44}=B_1^2B_2^2+|D_{12}|^2+|\bar D_{12}|^2+|C_1|^2|C_2|^2 -B_1^2|C_2|^2-B_2^2|C_1|^2-2B_1B_2|D_{12}|^2 -2|D_{12}|^2|\bar D_{12}|^2 \nonumber \\
& &\hspace{8mm}-4B_1{\rm Re}\left[C_2\bar D_{12}^*D_{12}^*\right] -4B_2{\rm Re}\left[C_1\bar D_{12}D_{12}^{*}\right]-2{\rm Re}\left[C_1C_2D_{12}^{*2}\right] -2{\rm Re}\left[C_1C_2^{*}\bar D_{12}^2\right].\nonumber
\end{eqnarray}

The considered noisy twin beams are characterized by the following
parameters~\cite{Perina1991Book,Arkhipov15}
\begin{eqnarray} % 20
 &B_1=B_{\rm p}+B_{\rm s}, \quad B_2=B_{\rm p}+B_{\rm i}, \quad D_{12}=i\sqrt{B_{\rm p}(B_{\rm p}+1)},&\nonumber \\
 &C_1 = C_2 = \bar{D}_{12}= 0&
\label{20}
\end{eqnarray}
where $B_{\rm p}$ is the mean photon-pair number and $ B_{\rm s}$
($ B_{\rm i} $) stands for the mean signal (idler) noise photon
number.

The transformation of a twin beam through the beam splitter can be
treated at the level of its covariance matrix $ {\bf A}_{\cal N}
$. The covariance matrix $ {\bf A}^{\rm out}_{\cal N} $
appropriate for the state at the output of a beam splitter with
transmissivity $ T $ is found as $ {\bf A}^{\rm out}_{\cal N} =
{\bf U}^{\dagger}{\bf A}^{\rm in}_{\cal N} {\bf U} $, where the
covariance matrix $ {\bf A}^{\rm in}_{\cal N} $ characterizes the
impinging twin beam and symbol ${\bf U}$ stands for the following
unitary matrix:
\begin{equation}\label{21}      % 21
 {\bf U}=\left(\begin{array}{cccc}
  \sqrt{T} & 0 &-\sqrt{R}\exp({i\phi})&0 \\
  0 & \sqrt{T}& 0 & -\sqrt{R}\exp({-i\phi})  \\
  \sqrt{R}\exp({-i\phi}) & 0 &\sqrt{T} & 0\\
  0 & \sqrt{R}\exp({i\phi}) & 0 & \sqrt{T}\end{array}\right),
\end{equation}
$R=1-T$. The phase $\phi$ occurring in Eq.~(\ref{21}) can be set
to zero without the loss of generality. The application of the
beam-splitter transformation (\ref{21}) to an input noisy twin
beam with parameters given in Eq.~(\ref{20}) leaves us with the
following two-mode Gaussian state:
\begin{eqnarray}   % 22
 B_1^{\rm out} &=& -T B_{\rm s} - B_{\rm p}- R B_{\rm i},\nonumber \\
 B_2^{\rm out} &=& -T B_{\rm i} - B_{\rm p}- R B_{\rm s},\nonumber \\
 C_1^{\rm out} = -C_2^{{\rm out}} &=& 2i\sqrt {TR}\sqrt{ B_{\rm p}(B_{\rm p}+1)},
  \nonumber \\
 D_{12}^{\rm out} &=& i(2T-1)\sqrt{ B_{\rm p}(B_{\rm p}+1)}, \nonumber \\
 \bar{D}_{12}^{\rm out} &=& \sqrt{TR}(B_{\rm s}- B_{\rm i}) .
\label{22}
\end{eqnarray}

Alternatively, we may derive explicit formulas for photon-number
distributions of both the impinging noisy twin beam and the state
at the output of the beam splitter. The photon-number distribution
$ p(n_1,n_2) $ of a noisy twin beam has been found in
Ref.~\cite{Arkhipov15}:
\begin{eqnarray}   % 23
 p(n_1,n_2) = \frac{1}{\tilde
   K}\sum_{m=0}^{{\rm min}(n_1,n_2)}
   \left( \begin{array}{c} n_1 \\ m \end{array} \right)
   \left( \begin{array}{c} n_2 \\ m \end{array} \right)
   \left( 1-\frac{\tilde B_1}{\tilde K} \right)^{n_2-m}
   \left( 1-\frac{\tilde B_2}{\tilde K} \right)^{n_1-m}
   \left(\frac{|D_{12}|}{\tilde K}\right)^{2m},
\end{eqnarray}
where $\tilde B_j=B_j+1$ for $j=1,2$ and $\tilde K = \tilde
B_1\tilde B_2-|D_{12}|^2 $. On the other hand, the photon-number
distribution $ p^{\rm out}(n_1,n_2) $ of the state at the
beam-splitter output is determined along the
formula~\cite{Kim2002}:
\begin{equation} % 24
 p^{\rm out}(n'_1,n'_2)=
 \sum_{n_1=0}^{\infty}\sum_{n_2=0}^{\infty} B_{n_1,n_2}^{n'_1,n'_2}
 p(n_1,n_2).
\label{24}
\end{equation}
The coefficients $ B_{n_1,n_2}^{n'_1,n'_2} $ in Eq.~(\ref{24}) are
defined as
\begin{eqnarray}  % 25
 B_{n_1,n_2}^{n'_1,n'_2} = \sum_{k_1=0}^{n_1} \sum_{k_2=0}^{n_2} (-1)^{n_1-k_1}
  \sqrt{R}^{n_1+n_2-k_1-k_2} \sqrt{T}^{k_1+k_2}  \frac{\sqrt{n_1!n_2!n'_1!n'_2!} }{k_1!(n_1-k_1)!k_2!(n_2-k_2)!}
  \delta_{n'_1,n_2+k_1-k_2}\delta_{n'_2,n_1-k_1+k_2}
\label{25}
\end{eqnarray}
and $ \delta $ means the Kronecker symbol.

The local non-classicality quantifiers $I_{\rm ncl}^{(j)}$, $
j=1,2 $ and entanglement quantifier $I_{\rm ent}$ introduced in
Ref.~\cite{Arkhipov2016b} have been found suitable as theoretical
characteristics for the analyzed two-mode Gaussian states. The
reason is that these quantifiers together form the global
non-classicality invariant $ I_{\rm ncl} $,
\begin{equation} % 26
 I_{\rm ncl} = I_{\rm ncl}^{(1)}+ I_{\rm ncl}^{(1)}+2I_{\rm ent},
\label{26}
\end{equation}
that does not change when any photon-number preserving unitary
transformation is applied. According to
Refs.~\cite{Arkhipov2016b,Arkhipov2016c} the local
non-classicality quantifier $I_{\rm ncl}^{(j)}$ for mode $ j $ is
given as:
\begin{equation} % 27
  I_{\rm ncl}^{(j)} = - B_j^2 + |C_j|^2.
\label{27}
\end{equation}
On the other hand, both three local and one global invariants of
the symmetrically-ordered covariance matrix $ {\bf A}_{\cal S} $
are needed to determine the entanglement quantifier $I_{\rm ent}$.
Details can be found in Ref.~\cite{Arkhipov2016c}.

\section*{Identification of non-classicality of twin beams}

We first consider the simplest case of a noiseless twin beam whose
only parameter is the mean photon-pair number $ B_{\rm p} $. Its
entanglement, which is responsible for its non-classicality, has
been theoretically analyzed in Ref.~\cite{Arkhipov15} where
negativity $ N $, which is an entanglement
measure~\cite{Plenio05}, has been derived as $ N = \sqrt{B_{\rm
p}(B_{\rm p}+1)}+B_{\rm p}$. Thus, the entanglement of a noiseless
twin beam increases with the photon-pair number $ B_{\rm p} $. In
our analysis, we consider the first five GNCCa $ E_{k_1,k_2}^W $
and $E_{k_1,k_2}^p$ that contain the intensity moments up to the
sixth order and the elements of photon-number distribution for up
to six photons. We note that the consideration of lower-order
intensity moments is natural as the experimental error increases
with the increasing order of intensity moments.

The GNCCa $ E^W $ and $ M^W $ given in Eqs.~(\ref{8}) and
(\ref{11}), respectively, and using intensity moments attain in
the case of a noiseless twin beam the forms:
\begin{eqnarray}   % 28-29
 E^W_{0,0} &=& -2B_{\rm p}, \nonumber \\
 E^W_{1,1} &=& -12B_{\rm p}^3-8B_{\rm p}^2,  \nonumber \\
 E^W_{2,2} &=& -240B_{\rm p}^5-288B_{\rm p}^4-72B_{\rm p}^3, \nonumber \\
 E^W_{0,1} &=&  -4B_{\rm p}^2, \nonumber \\
 E^W_{0,2} &=&  -12B_{\rm p}^3+4B_{\rm p}^2,
\label{28} \\
 M^W &=& -4B_{\rm p}^3-B_{\rm p}^2.
\label{29}
\end{eqnarray}
On the other hand, their counterparts $ E^p $ and $ M^p $
involving the elements of photon-number distribution and written
in Eqs.~(\ref{9}) and (\ref{11}), respectively, are obtained as:
\begin{eqnarray}   % 30-31
 E^p_{0,0} &=& -{2B_{\rm p}}/{(B_{\rm p}+1)}, \nonumber \\
 E^p_{1,1} &=& -{8B_{\rm p}^2}/{(B_{\rm p}+1)^2}, \nonumber \\
 E^p_{2,2} &=& -{72B_{\rm p}^3}/{(B_{\rm p}+1)^3}, \nonumber \\
 E^p_{0,1} &=& 0, \nonumber \\
 E^p_{0,2} &=& 4B_{\rm p}^2/{(B_{\rm p}+1)^2},
\label{30} \\
 M_p &=& -\frac{B_{\rm p}^2}{(B_{\rm p}+1)^2}.
\label{31}
\end{eqnarray}

\begin{figure}  %Figure 1
 \centerline{\includegraphics[width=0.85\textwidth]{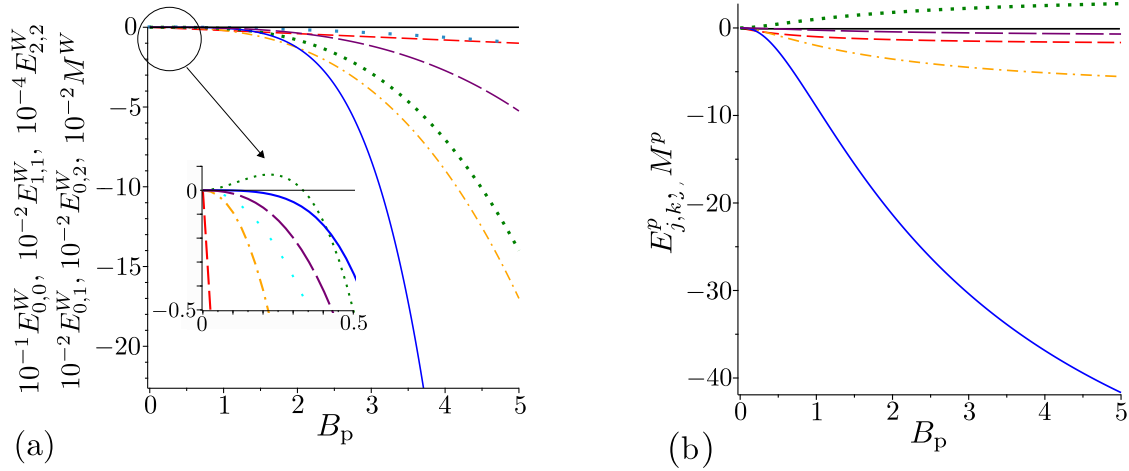}}
 \caption{(a) GNCCa $ E_{0,0}^W $ (red dashed curve), $E_{1,1}^W$ (orange dash-dotted curve),
  $ E_{2,2}^W$ (blue solid curve), $E_{0,1}^W$ (cyan long dotted  curve), $E_{0,2}^W$ (green dotted
 curve), and $M^W$ (purple long dashed curve) and (b) GNCCa $ E_{0,0}^p $ (red dashed curve), $E_{1,1}^p$
 (orange dash-dotted curve), $E_{2,2}^p$ (blue solid line), $E_{0,2}^p$ (green dotted curve),
 and $M^p$ (purple long dashed curve) as they depend on photon-pair
 number $ B_{\rm p} $ for noiseless twin beams.}
\label{fig1}
\end{figure}

Mutual comparison of the formulas for GNCCa written in
Eqs.~(\ref{28})---(\ref{31}) reveals qualitatively different
behavior of these GNCCa for greater photon-pair numbers $ B_{\rm
p} $ (see Fig.~\ref{fig1}). Whereas the GNCCa based on intensity
moments tend to go to minus infinity, the GNCCa using the elements
of photon-number distributions reach finite values for $ B_{\rm p}
\rightarrow \infty $.

The GNCCa $ E_{k,k}^W $, $ k=0,1, \ldots $, $ E_{0,1}^W $, and $
M^W $ as well as the GNCCa $ E_{k,k}^p $, $ k=0,1, \ldots $, and $
M^p $ are entanglement monotones, since their absolute values
increase with the increasing photon-pair number $B_{\rm p}$. As
such, all of them (with the inverted sign) can be chosen as a
suitable non-classicality identifier for any noiseless twin beam.
We note that this ability to reveal the non-classicality is
preserved for non-ideal detection with a finite detection
efficiency.

On the other hand, the GNCC $ E_{0,2}^W $ can be successfully
applied only for $ B_{\rm p} \in (1/3,\infty)$ and the GNCC $
E_{0,2}^p $ even attains positive values for any value of $ B_{\rm
p} $. A more general analysis has shown that the GNCCa $
E_{k_1,k_2}^W $ for $ |k_1 - k_2|>1 $ reveal non-classicality only
for more intense twin beams and the GNCCa $ E_{k_1,k_2}^p $ for $
k_1 \neq k_2 $ cannot indicate non-classicality at all.

\begin{figure} %Figure 2
\centerline{\includegraphics[width=0.75\textwidth]{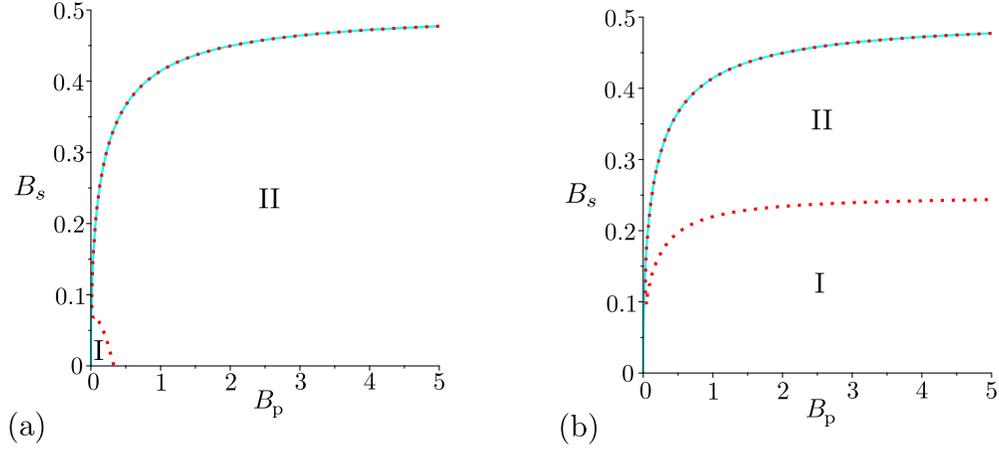}}
\caption{Non-classicality phase diagrams for noisy twin beams with
 balanced noise: (a) GNCCa
 $E_{0,0}^W$, $E_{1,1}^W$, $E_{2,2}^W$ (coinciding cyan solid curves),
 and $E_{0,2}^W $ (red dotted curve) and (b) GNCCa $E_{0,0}^p$,
 $E_{1,1}^p$, $E_{2,2}^p$ (coinciding cyan solid curves), and $E_{0,2}^p$ (red dashed curve)
 in space ($ B_{\rm s}$,$ B_{\rm p}$).
 For comparison, phase diagram of theoretical entanglement quantifier $ I_{\rm ent} $
 is plotted in (a) and (b) by cyan solid curve.
 The GNCCa $E_{0,2}^W$ and $E_{0,2}^p$ express non-classicality
 only in region II, the remaining GNCCa are negative in regions I
 and II.}
\label{fig2}
\end{figure}

Now we pay attention to noisy twin beams, first considering the
beams with balanced noise for which the signal and idler mean
noise photon numbers equal ($B_{\rm s}=B_{\rm i}$). The GNCCa $
E_{k,k}^W $, $ k=0,1,\ldots $, and $ M^W $ and the GNCCa $
E_{k,k}^p $, $ k=0,1,\ldots $, and $ M^p $ still fully identify
non-classicality of such noisy twin beams, that is, however,
observed only for twin beams with smaller amount of the noise (see
Fig.~\ref{fig2}). As it has been found in Ref.~\cite{Arkhipov15},
only the twin beams with $B_{\rm s}=B_{\rm i} < \sqrt{B_{\rm
p}(B_{\rm p}+1)}-B_{\rm p}$ are nonclassical.

Interestingly, the GNCCa $ E_{0,2}^W$ and $E_{0,2}^p$ indicate
non-classicality of noisy twin beams with photon-pair numbers $
B_{\rm p} $ for which they failed in the case of noiseless twin
beams. This occurs for the GNCC $E_{0,2}^W$ in region $ B_{\rm
p}\in(0,1/3)$ for the noisy twin beams with $B_{\rm s} = B_{\rm i}
\in ( [\sqrt{B_{\rm p}(B_{\rm p}+1)}-B_{\rm p}]/2,\sqrt{B_{\rm
p}(B_{\rm p}+1)}-B_{\rm p} )$ [region II in Fig.~\ref{fig2}(a)].
Similarly, the GNCC $E_{0,2}^p$, that is nonnegative for noiseless
twin beams, is negative for the noisy twin beams with $ B_{\rm s}
= B_{\rm i} \in ( [\sqrt{4B_{\rm p}(B_{\rm p}+1)+2\sqrt{B_{\rm
p}(B_{\rm p}+1)}+1}-2B_{\rm p}-1]/2,\sqrt{B_{\rm p}(B_{\rm
p}+1)}-B_{\rm p} )$ [region II in Fig.~\ref{fig2}(b)].

\begin{figure} %Figure 3
 \centerline{\includegraphics[width=0.98\textwidth]{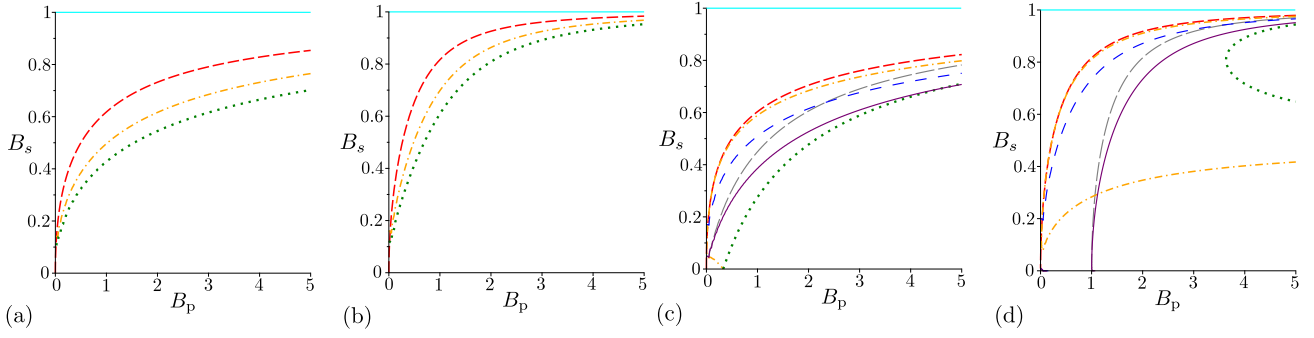}}
 \caption{Non-classicality phase diagrams for noisy twin beams with unbalanced noise ($B_{\rm i}=0$):
 (a) GNCCa $E_{0,0}^W$ (red dashed curve), $E_{1,1}^W$ (orange dash-dotted curves),
  $E_{2,2}^W$ (green dotted curve), and $M^W$ (cyan solid curve),
 (b) GNCCa $E_{0,0}^p$ (red dashed curve), $E_{1,1}^p$ (orange
  dash-dotted curve), $E_{2,2}^p$ (green dotted curve), and $ M^p$
  (cyan solid curve),
 (c) $E_{1,0}^W$ (red dashed curve),
  $E_{0,1}^W$ (grey long dashed curve), $E_{2,0}^W$ (orange
  dash-dotted curves), $E_{0,2}^W$ (green dotted curve), $E_{2,1}^W$
  (blue space dashed curve), and $E_{1,2}^W$ (purple solid curve)
  and (d) GNCCa $E_{1,0}^p$ (red dashed curve), $E_{0,1}^p$ (grey
  long dashed curve), $E_{2,0}^p$ (orange dash-dotted curves),
  $E_{0,2}^p$ (green dotted curves), $E_{2,1}^p$ (blue space dashed
  curve), and $E_{1,2}^p$ (purple solid curve) in space ($ B_{\rm
  s}$,$ B_{\rm p}$). For comparison, phase diagram of entanglement
  quantifier $ I_{\rm ent} $ is plotted by cyan solid curve in
  (a)---(d). For GNCCa $E_{2,0}^W$ plotted in (c) and $E_{2,0}^p$ [$E_{0,2}^p$]
  drawn in (d), the non-classicality region lies between the lower and
  upper orange dash-dotted [green dotted] curves. For the other
  GNCCa, the non-classicality region is below the corresponding
  curves.}
\label{fig3}
\end{figure}

Finally, we analyze the performance of the above discussed GNCCa
when they are applied to the noisy twin beams with unbalanced
noise. We assume that the noise is present only in the signal
field ($ B_{\rm s}\neq 0 $, $ B_{\rm i}= 0 $). Such twin beams
have been theoretically investigated in Ref.~\cite{Arkhipov15}
with the conclusion that only the twin beams with $ B_{\rm s}<1 $
and arbitrary $ B_{\rm p} $ exhibit non-classicality. The GNCCa $
E_{k_1,k_2}^W$ and $ E_{k_1,k_2}^p$ reveal non-classicality only
for some noisy twin beams, especially those with smaller amount of
the noise [see Fig.~\ref{fig3}]. The GNCCa $ E_{k_1,k_2}^W$ and $
E_{k_1,k_2}^p$ with $ k_1 > k_2 $ are more sensitive to
non-classicality as they include higher-order signal-field
intensity moments and the elements of the photon-number
distribution for greater signal photon numbers, respectively. This
is the consequence of the noise present in the signal field.
Contrary to this, the GNCCa $ M^W $ and $ M^p $ are able to
indicate non-classicality for all noisy twin beams, as documented
in Figs.~\ref{fig3}(a) and \ref{fig3}(b). This means that the
GNCCa $ M^W $ and $ M^p $ allow to reveal non-classicality of all
analyzed twin beams.

\section*{Identification of non-classicality of two-mode states beyond a beam splitter}

In this section, we address two-mode states that occur at the
output ports of a beam splitter with transmissivity $ T $ assuming
an input noisy twin beam. We note that in the boundary cases $ T=0
$ and $ T=1$ an input noisy twin beam is just transformed to the
beam-splitter output without any modification. On the other hand,
the balanced beam splitter with $ T=1/2 $ is optimal for the
generation of squeezed light in both output
ports~\cite{Paris97,Arkhipov2016b,Arkhipov2016c}. In general, an
arbitrary beam splitter has the potential to generate states that
may exhibit both local non-classicality and entanglement.

\begin{figure}  %Figure 4
 \centerline{\includegraphics[width=0.85\textwidth]{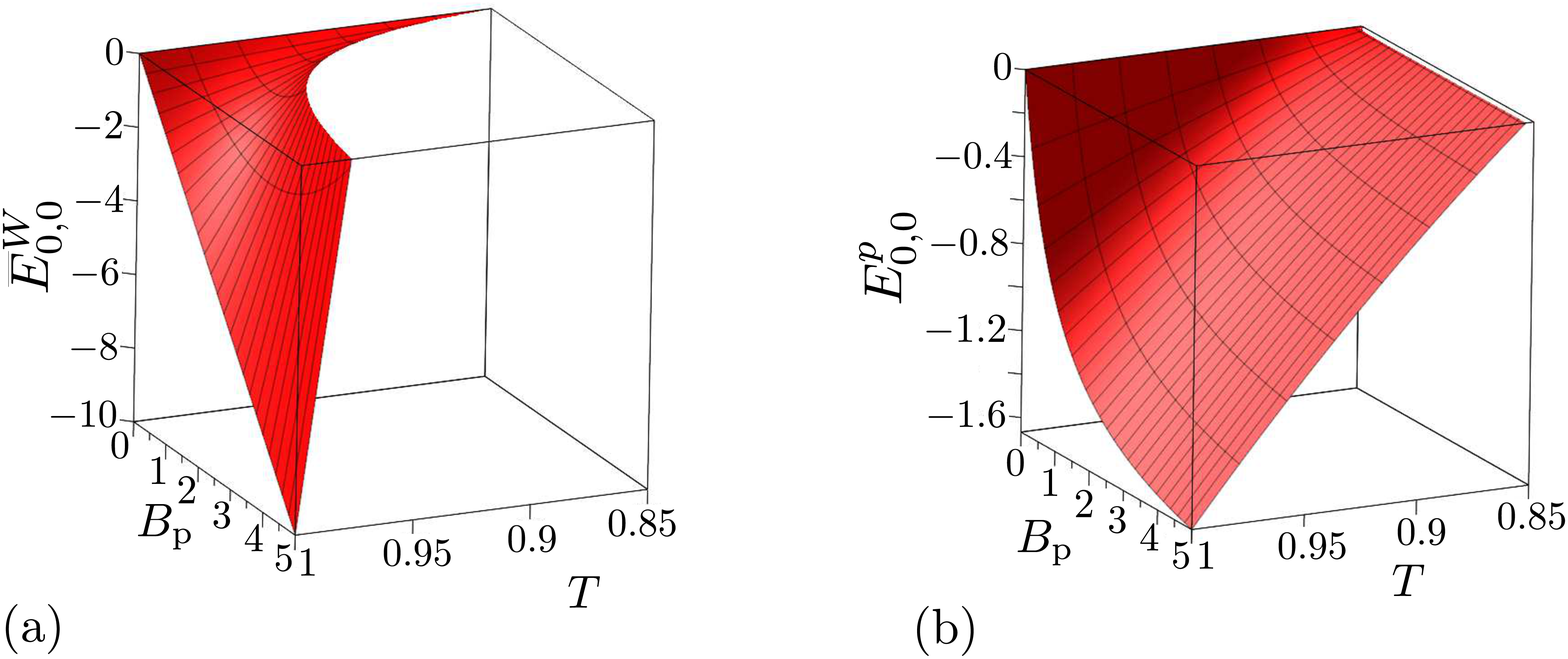}}
 \caption{(a) GNCC $E_{0,0}^W$ and (b) GNCC $E_{0,0}^p$ as
  functions of photon-pair number $B_{\rm p}$ and beam-splitter
  transmissivity $T$ for noiseless two-mode states beyond the beam splitter.}
\label{fig4}
\end{figure}

\begin{figure}  %Figure 5
 \centerline{\includegraphics[width=0.75\textwidth]{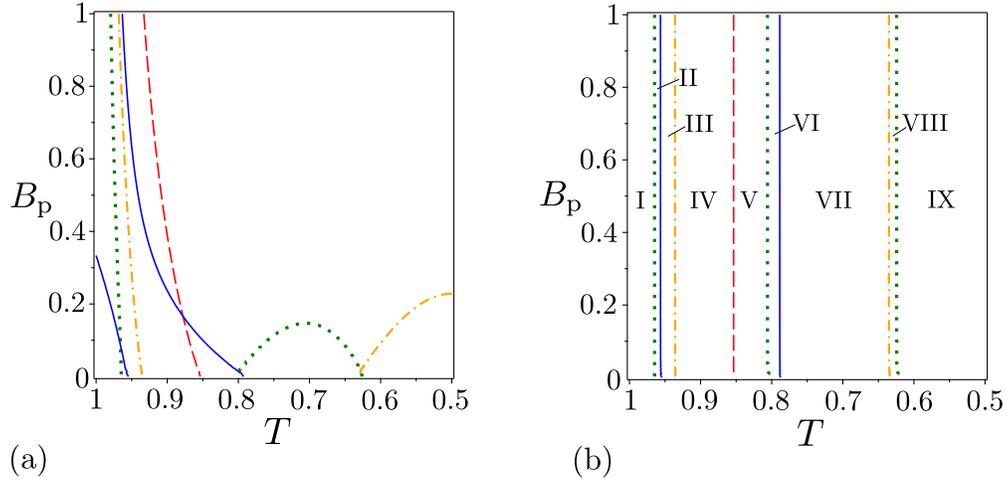}}
 \caption{Non-classicality phase diagrams for noiseless two-mode states beyond the beam splitter:
 (a) GNCCa $E_{0,0}^W$ (red dashed curve), $E_{1,1}^W$ (orange dash-dotted
  curve), $E_{2,2}^W$ (green dotted curve), $E_{2,0}^W$ (blue solid
  curve), and $M^W$ (red dashed curve) [non-classicality regions are found below the corresponding
  curves, except that for the GNCC $E_{2,0}^W$ occurring between the blue
  curves] and
 (b) $E_{0,0}^p$ (red dashed line), $E_{1,1}^p$ (orange dash-dotted
  lines), $E_{2,2}^p$ (green dotted lines), $E_{2,0}^p$ (blue solid
  lines), and $M^p$ (red dashed line) [non-classicality regions for different GNCCa are identified in
  Table~\ref{tab1}] in space ($ B_{\rm p}$,$ T $).}
\label{fig5}
\end{figure}

Similarly as in the previous section, we analyze the noiseless
states first. It is interesting that the GNCCa $ E_{k_1,k_2}^p $
and $ M^p $ involving the elements of photon-number distributions
factorize as functions of transmissivity $ T $ and photon-pair
number $ B_{\rm p} $, contrary to the GNCCa $ E_{k_1,k_2}^W $ and
$ M^W $ based on intensity moments (see the graphs for $ E_{0,0}^W
$ and $ E_{0,0}^p $ in Fig.~\ref{fig4} and also the
non-classicality phase diagrams in Fig.~\ref{fig5}).
In particular, the following formulas can be derived:
\begin{eqnarray}   % 32-33
 E_{0,0}^p &=& -(1-8TR)\frac{2B_{\rm p}}{(B_{\rm p}+1)}, \nonumber \\
 E_{1,1}^p &=& -(72T^2R^2-21TR+1)\frac{8B_{\rm p}^2}{(B_{\rm p}+1)^2}, \nonumber \\
 E_{2,2}^p &=& -(800T^3R^3+340T^2R^2-40TR+1)\frac{72B_{\rm p}^3}{(B_{\rm p}+1)^3}. \nonumber \\
 E_{0,2}^p &=& (144T^2R^2-30TR+1)\frac{4B_{\rm p}^2}{(B_{\rm p}+1)^2},
\label{32} \\
 M^p &=& -(1-8TR)\frac{B_{\rm p}^2}{(B_{\rm p}+1)^2}.
\label{33}
\end{eqnarray}
Due to the factorization, the corresponding non-classicality
regions do not depend on photon-pair number $B_{\rm p}$. Detailed
analysis of the non-classicality regions whose results are
summarized in Table~\ref{tab1} has shown that the GNCCa $ E_{0,0}^p
$, $ E_{1,1}^p $, $ E_{2,2}^p $, and $ E_{0,2}^p $ considered
together allow to reveal non-classicality of an arbitrary
noiseless two-mode state beyond the beam splitter with
transmissivity $T\in[1/2,1]$ (and also $T\in[0,1/2]$ due to the
symmetry reasons). As the phase diagram plotted in
Fig.~\ref{fig5}(a) documents, the set of GNCCa $ E_{0,0}^W $, $
E_{1,1}^W $, $ E_{2,2}^W $, and $ E_{0,2}^W $ involving intensity
moments allows to detect non-classicality of all considered states
only for small photon-pair numbers $ B_{\rm p} $. As evident from
Figs.~\ref{fig4}(a) and \ref{fig5}(a), these GNCCa lose their
ability to reveal non-classicality with the increasing photon-pair
number $ B_{\rm p} $.
%\begin{widetext}
\begin{center}
\begin{table*}[t!]  % Table I
\setlength\extrarowheight{11pt}
 \begin{center}
 \begin{tabular}{| c | c |c| }
  \hline GNCC & Non-classicality region(s) & Corresponding areas in Fig.~5(b) \\
  \hline $E_{0,0}^p$ & $T\in( [1+1/\sqrt{2}]/2, 1 ]$ & I,II,III,IV \\
  \hline $E_{1,1}^p$ & $T\in [ 1/2,[1+\sqrt{15-3\sqrt{17}}/6]/2 )\cup ( [1+\sqrt{15+3\sqrt{17}}/6]/2, 1 ]$
    & I,II,III,VIII,IX  \\
  \hline $E_{2,2}^p$ & $T \in (\approx 0.624,\approx 0.806) \cup (\approx 0.965, 1 ]$ & I, VI, VII, VIII \\
  \hline $E_{0,2}^p$ & $T\in ( [1+1/\sqrt{3}]/2, [1+\sqrt{30}/6]/2 )$
    & III, IV, V, VI  \\
  \hline
 \end{tabular}
 \end{center}
 \caption{Non-classicality regions of GNCCa $ E_{0,0}^p $, $ E_{1,1}^p $, $ E_{2,2}^p $, and $ E_{0,2}^p $ defined on
  the beam-splitter transmissivity axis $ T $ for noiseless two-mode states beyond the beam splitter.}
\label{tab1}
\end{table*}
\end{center}
%\end{widetext}

The GNCCa $ E_{0,0}^p $, $ E_{2,2}^p $ and $ E_{0,2}^p $ shown in
the phase diagram in Fig.~\ref{fig5}(b) detect
entanglement~\cite{PerinaJr2017a} and so they lose their ability
to reveal global non-classicality as $ T $ approaches 1/2. The
reason is that the entanglement of the considered states is
becoming weaker as $ T $ goes to 1/2 and the state is separable
for $ T=1/2 $. On the other hand, the GNCC $ E_{1,1}^p $ safely
indicates global non-classicality in the region around $ T=1/2 $.
This is understood by the fact that the GNCC $ E_{1,1}^p $ is able
to reveal also local non-classicality [$ E_{1,1}^p $ given in
Eq.~(\ref{9}) and $ R_{1,1}^p $ defined in Eq.~(\ref{12}) coincide
for separable symmetric ($ 1 \leftrightarrow 2 $) states]. We note
that the vanishing entanglement in the vicinity of $T=1/2$ can
only be identified by the GNCCa $E_{2k,2k}^p$, $ k=1,2,\ldots $.
The greater the number $ k $ is the two-mode entangled states
generated with $ T $ closer to 1/2 can be revealed. However, this
requires the determination of photon-number distributions for
greater photon numbers~\cite{Harder2016}.

\begin{figure} %Figure6
\centerline{\includegraphics[width=0.75\textwidth]{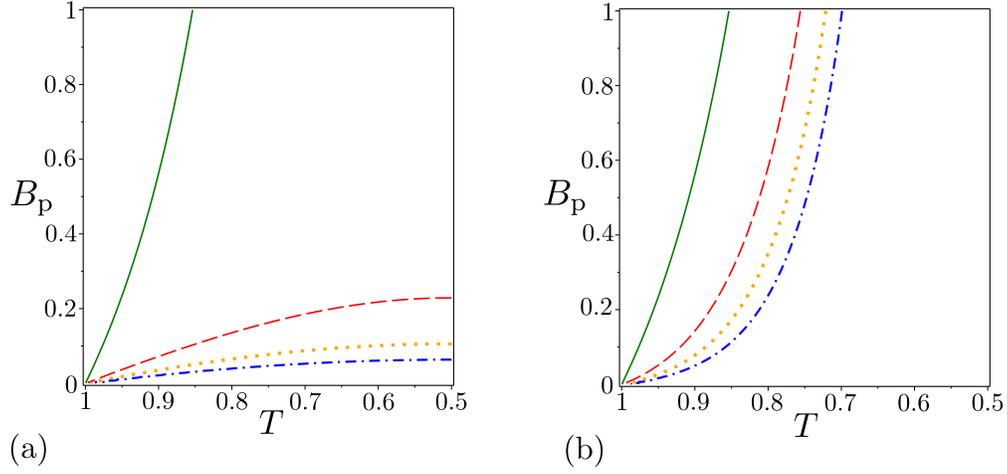}}
\caption{Non-classicality phase diagrams for noiseless two-mode
 states beyond the beam splitter: (a) $R_{2,2}^W$ (red
 dashed curve), $R_{4,4}^W$ (orange dotted curve) and $R_{6,6}^W$ (blue
 dash-dotted curve) and (b) $R_{2,2}^p$ (red dashed curve), $R_{4,4}^p$ (orange dotted curve) and
 $R_{6,6}^p$ (blue dash-dotted curve) in space ($ B_{\rm p}$,$ T $). For comparison,
 phase diagram of local non-classicality
 quantifier $I_{\rm ncl}^{(1)}=I_{\rm ncl}^{(2)}$ is plotted by green solid
 curve. Non-classicality regions extend from line $ B_{\rm p} = 0
 $ up to the corresponding curves drawn in the diagrams.}
\label{fig6}
\end{figure}

The striking feature of two-mode states beyond the beam splitter
is the ability to exhibit local non-classicality. This originates
in the bunching effect of photons in a photon pair at a beam
splitter. Ideally, two non-distinguishable photons impinging on a
balanced beam splitter leave the beam splitter at the same output
port. Thus, the original twin beam partly loses its entanglement
as it propagates through the beam splitter, but its constituting
parts can gain their local non-classicalitites, as quantified by
relation~(\ref{26}) for the global non-classicality invariant $
I_{\rm ncl} $. As local non-classicality arises from pairing of
photons, only the LNCCa $ R_{2k,2k}^W $ and $ R_{2k,2k}^p $, $
k=1,2,\ldots $, allow for detecting local non-classicality. The
phase diagram for the local non-classicality quantifier $ I_{\rm
nlc}^{(1)} $ in Fig.~6 shows that the majority of the considered
states with smaller photon-pair numbers $ B_{\rm p} $ exhibit
local non-classicality. However, the analyzed LNCCa $
R_{2k,2k}^{W} $ and $ R_{2k,2k}^{p} $ for $ k=1,2 $ and 3, whose
phase diagrams are also included in Fig.~6, identify local
non-classicality only in some states. As the identifiable states
are in the area around $ T=1/2 $ they are apparently endowed with
stronger local non-classicality. As documented in Fig.~\ref{fig6},
the LNCCa $ R_{2k,2k}^{W} $, $ k=1,2,\ldots $, based on intensity
moments are applicable only to weak two-mode fields and they lose
their power with the increasing index $ k $. Also the LNCCa $
R_{2k,2k}^{p} $, $ k=1,2,\ldots $, determined from the elements of
photon-number distribution gradually lose their power with the
increasing index $ k $, but they are suitable for indicating local
non-classicality in more intense two-mode fields [see
Fig.~\ref{fig6}(b)]. The LNCC $ R_{2,2}^{p} $ is the most powerful
among the studied LNCCa and, assuming the beam splitter with fixed
transmissivity $ T $, it allows to reveal the local
non-classicality of two-mode states with photon-pair numbers $
B_{\rm p} $ lower than
\begin{equation}   % 34
 B_{\rm p}=\frac{4TR-1+\sqrt{4TR\left[4TR-(7+\sqrt{33})\right]+1}}{2(2T-1)^2}.
\label{34}
\end{equation}

\begin{figure}  % Figure 7
\centerline{\centerline{\includegraphics[width=\textwidth]{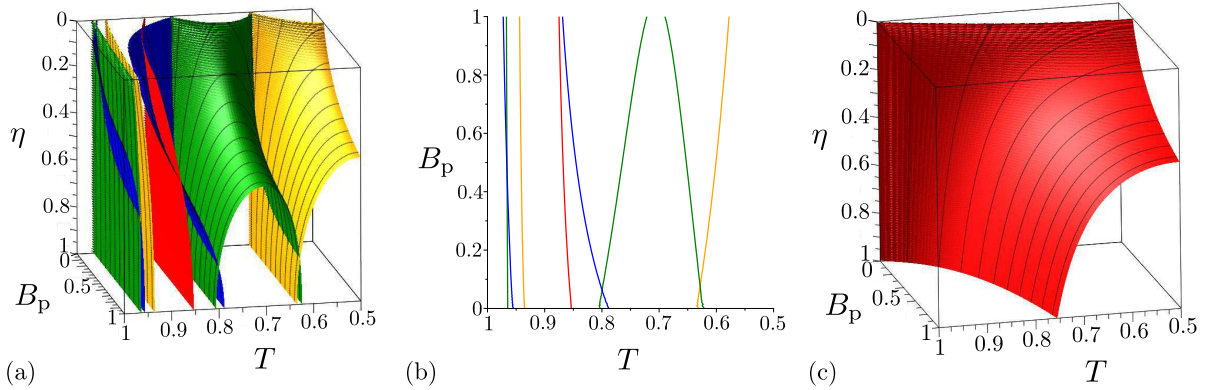}}}
\caption{Non-classicality phase diagrams for noiseless two-mode
 states beyond the beam splitter: (a) GNCCa $E_{0,0}^p$ (red
 surface), $E_{1,1}^p$ (orange contour surface), $E_{2,2}^p$ (green
 contour surface), $E_{2,0}^p$ (blue contour surface), and $M^p$ (red
 surface), (b) plane $\eta=0.5$ in the phase diagrams plotted in (a) and
 (c) LNCC $R_{2,2}^p$ in space ($ B_{\rm p}$,$ T $,$ \eta $).
 We note that the planes $\eta=0$ and $\eta=1$ in the phase diagrams in (a)
 [(c)] are plotted in Figs.~\ref{fig5}(a) and \ref{fig5}(b) [Figs.~\ref{fig6}(a) and \ref{fig6}(b) by red dashed curves],
 respectively. Non-classicality regions occur below the corresponding surfaces, only that for the GNCC $E_{2,0}^p$ in (a) is
 found between blue contour surfaces.}
\label{fig7}
\end{figure}

In real experimental identification of both global and local
non-classicalities finite detection efficiencies are an important
issue. The GNCCa as well as LNCCa based on intensity moments are
not sensitive to detection efficiency $ \eta $ because the moments
in these criteria are only synchronously rescaled with appropriate
powers of efficiency $ \eta $. Contrary to this, the GNCCa and
LNCCa containing the elements of photon-number distribution suffer
from the finite detection efficiency $ \eta $. Gradual loss of the
power to resolve nonclassical states with decreasing detection
efficiency $ \eta $ is documented in Fig.~\ref{fig7}(a) for the
GNCCa $ E_{0,0}^p $, $ E_{1,1}^p $, $ E_{2,2}^p $, $ E_{2,0}^p $,
and $ M_p $ and in Fig.~\ref{fig7}(c) for the LNCC $ R_{2,2}^p $.
Except for the GNCC $ E_{2,0}^p $, the set of nonclassical
two-mode states identified by the other analyzed GNCCa and LNCCa
only diminishes with decreasing detection efficiency $ \eta $. For
the GNCC $ E_{2,0}^p $, nonclassical two-mode states with
decreasing photon-pair numbers $ B_{\rm p} $ are gradually
identified as the detection efficiency $ \eta $ decreases [compare
the corresponding phase diagrams in Figs.~\ref{fig5}(a) and
\ref{fig5}(b)].

In the limit $ \eta \rightarrow 0 $, the phase diagrams of the
GNCCa and LNCCa based on the elements of photon-number
distribution coincide with those written for intensity moments.
This property has its origin in the form of the Mandel
photodetection formula that provides the following relation for
small detection efficiencies $ \eta $:
\begin{equation}  % 35
  n_1! n_2! p(n_1,n_2) \approx \eta^{n_1+n_2} \langle W_1^{n_1}
  W_2^{n_2} \rangle .
\label{35}
\end{equation}
The process of gradual loss of the ability to detect
non-classicality with decreasing detection efficiency $ \eta $ can
be treated even analytically for individual GNCCa and LNCCa. For
example, we have for the GNCCa $ E_{0,0}^p $ and $ E_{0,0}^W $
\begin{equation}   % 36
 E_{0,0}^p = K [E_{0,0}^W -2\eta(2-\eta)B_{\rm p}^2 ]
\label{36}
\end{equation}
and $K= \eta^2/[1+\eta(2-\eta)B_{\rm p}]^{2}$ is a positive
constant. We have $ E_{0,0}^p = \eta^2 E_{0,0}^W $ for $ \eta $
approaching 0.

\begin{figure}  %Figure 8
\centerline{\includegraphics[width=\textwidth]{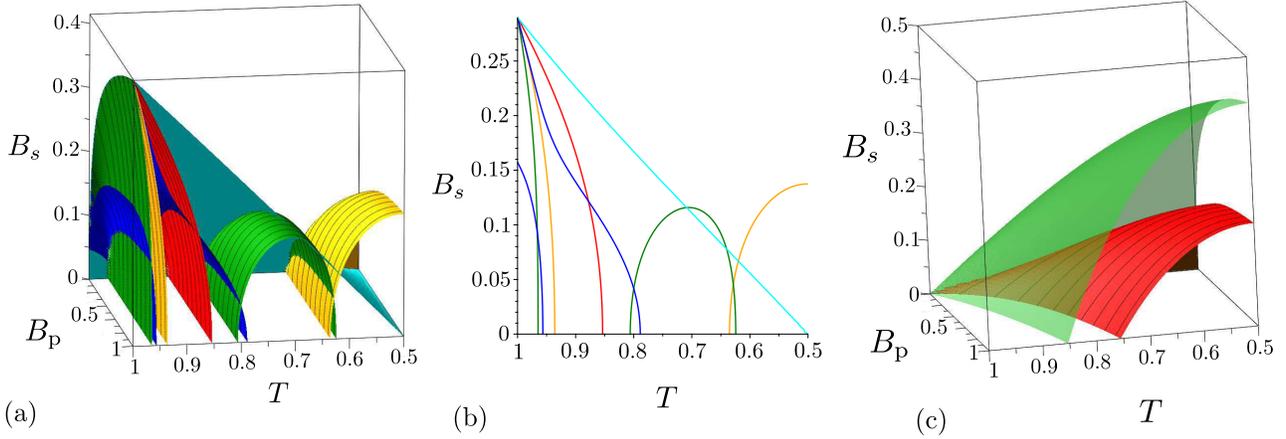}}
\caption{ Non-classicality phase diagrams for two-mode states
 beyond the beam splitter with input noisy twin beams with balanced
 noise $B_{\rm s}=B_{\rm i}$: (a) GNCCa
 $ E_{0,0}^p$ (red contour surface), $E_{1,1}^p$ (orange contour
 surface), $E_{2,2}^p$ (green contour surface), $E_{2,0}^p$ (blue
 contour surface), $ M^p$ (red contour surface)
 and entanglement quantifier $I_{\rm ent}$ (cyan surface),
 (b) plane $B_{\rm p}=0.2$ in the phase diagrams plotted in (a)
 and (c) LNCC $ R_{2,2}^p $ (red contour surface) and local non-classicality
 quantifiers $I_{\rm ncl}^{(1)}=I_{\rm ncl}^{(2)}$ (green surface)
 in space ($ B_{\rm p} $,$ T $,$ B_{\rm s} $). Non-classicality regions occur in (a) and (c) below the corresponding
 surfaces, except for the GNCC $E_{2,0}^p$ in (a) for which the non-classicality region is surrounded by blue contour surfaces.}
\label{fig8}
\end{figure}

Two-mode states occurring beyond the beam splitter with an
impinging noisy twin beam can be both locally nonclassical and
entangled. However, the numbers $ B_{\rm s} = B_{\rm i} $ of noise
photons cannot exceed the value $\sqrt{\B(\B+1)}-\B$ for twin
beams with balanced noise. It holds that the entangled two-mode
states are generated in the areas around $ T =1 $ and $ T=0 $ when
the input noise cannot be neglected. With the increasing numbers $
B_{\rm s} = B_{\rm i} $ of noise photons the areas containing
entangled states shrink towards the points $ T =1 $ and $ T=0 $
[see Fig.~\ref{fig8}(a)]. On the other hand, two-mode states
exhibiting local non-classicality occur in the area around $ T=1/2
$. With the increasing numbers $ B_{\rm s} = B_{\rm i} $ of noise
photons this area diminishes [see Fig.~\ref{fig8}(c)].
\begin{figure}  % Figure 9
\centerline{\includegraphics[width=\textwidth]{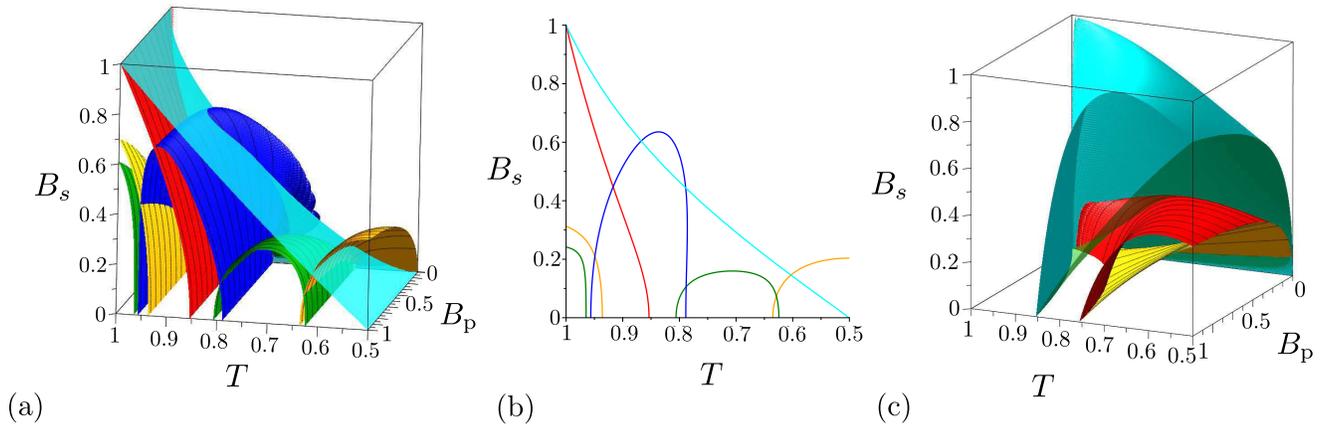}}
\caption{ Non-classicality phase diagrams for two-mode states
 beyond the beam splitter with input noisy twin beams with unbalanced
 noise: (a) GNCCa $E_{1,1}^p$
 (orange contour surface), $E_{2,2}^p$ (green contour surface),
 $E_{0,2}^p$ (blue contour surface), $M^p$ (red contour
 surface), and entanglement quantifier $I_{\rm ent}$ (cyan
 surface), (b) plane $B_{\rm p}=0.2$ in the phase diagrams plotted in (a)
 and (c) LNCCa $R_{2,2}^{p_1}$
(orange contour surface), $R_{2,2}^{p_2}$ (red contour surface)
 and local non-classicality quantifiers $I_{\rm ncl}^{(1)}$ (green surface) and
 $I_{\rm ncl}^{(2)}$ (cyan surface) in space ($ B_{\rm p} $,$ T $,$ B_{\rm s} $).
 Non-classicality regions in (a) and (c) occur below the corresponding
 surfaces.}
\label{fig9}
\end{figure}
Neither the entangled two-mode states nor the locally-nonclassical
two-mode states can be completely identified by the analyzed GNCCa
and LNCCa. For example, the LNCC $ R_{2,2}^p $, that performs the
best, identifies local non-classicality only for two-mode states
with the numbers $ B_{\rm s} = B_{\rm i} $ of noise photons below
the value
\begin{equation} % 37
 B_{\rm s} = B_{\rm i} =\frac{1}{2}\Bigg[\sqrt{16TR\B(\B+1)+4\sqrt{\sqrt{33}-5}\sqrt{TR\B(\B+1)}+1 }
  -2\B-1\Bigg].
\label{37}
\end{equation}

When the noisy twin beams with unbalanced noise ($ B_{\rm s} \neq
0 $, $ B_{\rm i}=0 $) are assumed at the beam splitter, the
generated two-mode states behave similarly as those analyzed in
the case of twin beams with balanced noise. To observe
non-classicality, the input noisy twin beams cannot contain more
than one noise signal photon on average ($ B_{\rm s}<1 $).
Contrary to the case with balanced noise, only the GNCC $ M^p $ is
powerful in identifying entanglement for two-mode states beyond
the beam splitter with transmissivity $ T $ close to 1 and 0 in
this case [see phase diagrams in Fig.~\ref{fig9}(a)]. Also the
GNCCa $ E_{0,2}^p $ and $ E_{2,0}^p $ perform differently. For
two-mode states generated for $ T \in (1/2,1] $ the GNCC $
E_{0,2}^p $ is more efficient and it allows to reveal the
entanglement of all states detectable by the GNCC $ E_{2,0}^p $.
Also the best performing LNCCa $ R_{2,2}^{p_1} $ and $
R_{2,2}^{p_2} $, that reveal local non-classicality, give
different results in different areas of their phase diagrams [see
Fig.~\ref{fig9}(c)]. Whereas the LNCC $ R_{2,2}^{p_1} $ is more
suitable for two-mode states generated for $ T\in [0,1/2) $, the
LNCC $ R_{2,2}^{p_2} $ is more powerful for indicating local
non-classicality of two-mode states reached for $ T\in (1/2,1] $,
as documented in Fig.~\ref{fig9}(c). We note that the mean number
$ B_{\rm s} $ of noise signal photons of the input twin beam is
divided into the output beam-splitter ports such that $ TB_{\rm s}
$ mean noise photons leave mode 1 and $ RB_{\rm s} $ mean noise
photons occur in mode 2.

\section*{Conclusions}

We have analyzed the performance of several local and global
non-classicality criteria written for intensity moments and
elements of photon-number distributions and applied to noisy twin
beams and other two-mode states derived from noisy twin beams by
using a beam splitter. It has been shown that the non-classicality
criteria based on the elements of photon-number distributions
exhibit in general better performance in revealing both local and
global non-classicalities compared to those containing intensity
moments. However, these criteria lose their power with decreasing
detection efficiencies and they give the same results as the
criteria based on intensity moments for low detection
efficiencies. Both types of criteria contain one criterion that
reveals the entanglement as one of the forms of global
non-classicality for all entangled noisy twin beams. Contrary to
this, not all locally and globally nonclassical two-mode states
occurring beyond the beam splitter are detectable by the analyzed
non-classicality criteria. However, simultaneous application of
several criteria gives a good chance for revealing possible
non-classicality of an unknown two-mode state generated beyond the
beam splitter.

\bibliographystyle{naturemag}
%\bibliography{arkhipov}

\begin{thebibliography}{10}
\expandafter\ifx\csname url\endcsname\relax
  \def\url#1{\texttt{#1}}\fi
\expandafter\ifx\csname urlprefix\endcsname\relax\def\urlprefix{URL }\fi
\providecommand{\bibinfo}[2]{#2}
\providecommand{\eprint}[2][]{\url{#2}}

\bibitem{NielsenBook}
\bibinfo{author}{Nielsen, M.~A.} \& \bibinfo{author}{Chuang, I.~L.}
\newblock \emph{\bibinfo{title}{Quantum Computation and Quantum Information}}
  (\bibinfo{publisher}{Cambridge University Press},
  \bibinfo{address}{Cambridge}, \bibinfo{year}{2000}).

\bibitem{Wilde2013}
\bibinfo{author}{Wilde, M.~M.}
\newblock \emph{\bibinfo{title}{{Q}uantum {I}nformation {T}heory}}
  (\bibinfo{publisher}{Cambridge Univ. Press, Cambridge},
  \bibinfo{year}{2013}).

\bibitem{Braunstein05_RMP}
\bibinfo{author}{Braunstein, S.~L.} \& \bibinfo{author}{van Loock, P.}
\newblock \bibinfo{title}{Quantum information with continuous variables}.
\newblock \emph{\bibinfo{journal}{Rev. Mod. Phys.}}
  \textbf{\bibinfo{volume}{77}}, \bibinfo{pages}{513--577}
  (\bibinfo{year}{2005}).

\bibitem{Weedbrook12}
\bibinfo{author}{Weedbrook, C.} \emph{et~al.}
\newblock \bibinfo{title}{Gaussian quantum information}.
\newblock \emph{\bibinfo{journal}{Rev. Mod. Phys.}}
  \textbf{\bibinfo{volume}{84}}, \bibinfo{pages}{621--669}
  (\bibinfo{year}{2012}).

\bibitem{MandelBook}
\bibinfo{author}{Mandel, L.} \& \bibinfo{author}{Wolf, E.}
\newblock \emph{\bibinfo{title}{Optical Coherence and Quantum Optics}}
  (\bibinfo{publisher}{Cambridge University Press},
  \bibinfo{address}{Cambridge}, \bibinfo{year}{1995}).

\bibitem{Perina1991Book}
\bibinfo{author}{Pe\v{r}ina, J.}
\newblock \emph{\bibinfo{title}{Quantum Statistics of Linear and Nonlinear
  Optical Phenomena}} (\bibinfo{publisher}{Kluwer, Dordrecht},
  \bibinfo{year}{1991}).

\bibitem{Kim2002}
\bibinfo{author}{Kim, M.~S.}, \bibinfo{author}{Son, W.},
  \bibinfo{author}{Bu\ifmmode~\check{z}\else \v{z}\fi{}ek, V.} \&
  \bibinfo{author}{Knight, P.~L.}
\newblock \bibinfo{title}{Entanglement by a beam splitter: Nonclassicality as a
  prerequisite for entanglement}.
\newblock \emph{\bibinfo{journal}{Phys. Rev. A}} \textbf{\bibinfo{volume}{65}},
  \bibinfo{pages}{032323} (\bibinfo{year}{2002}).

\bibitem{Lvovsky2009}
\bibinfo{author}{{A. I. Lvovsky and M. G. Raymer}}.
\newblock \bibinfo{title}{Continuous-variable optical quantum state
  tomography}.
\newblock \emph{\bibinfo{journal}{Rev. Mod. Phys.}}
  \textbf{\bibinfo{volume}{81}}, \bibinfo{pages}{299} (\bibinfo{year}{2009}).

\bibitem{Shchukin2006}
\bibinfo{author}{Shchukin, E.} \& \bibinfo{author}{Vogel, W.}
\newblock \bibinfo{title}{Universal measurement of quantum correlations of
  radiation}.
\newblock \emph{\bibinfo{journal}{Phys. Rev. Lett.}}
  \textbf{\bibinfo{volume}{96}}, \bibinfo{pages}{200403}
  (\bibinfo{year}{2006}).

\bibitem{Sperling2012a}
\bibinfo{author}{Sperling, J.}, \bibinfo{author}{Vogel, W.} \&
  \bibinfo{author}{Agarwal, G.~S.}
\newblock \bibinfo{title}{Sub-binomial light}.
\newblock \emph{\bibinfo{journal}{Phys. Rev. Lett.}}
  \textbf{\bibinfo{volume}{109}}, \bibinfo{pages}{093601}
  (\bibinfo{year}{2012}).

\bibitem{Haderka2005a}
\bibinfo{author}{Haderka, O.}, \bibinfo{author}{{Pe\v{r}ina~Jr.}, J.},
  \bibinfo{author}{Hamar, M.} \& \bibinfo{author}{Pe\v{r}ina, J.}
\newblock \bibinfo{title}{Direct measurement and reconstruction of nonclassical
  features of twin beams generated in spontaneous parametric down-conversion}.
\newblock \emph{\bibinfo{journal}{Phys. Rev. A}} \textbf{\bibinfo{volume}{71}},
  \bibinfo{pages}{033815} (\bibinfo{year}{2005}).

\bibitem{PerinaJr2013a}
\bibinfo{author}{{Pe\v{r}ina~Jr.}, J.}, \bibinfo{author}{Haderka, O.},
  \bibinfo{author}{Mich\'{a}lek, V.} \& \bibinfo{author}{Hamar, M.}
\newblock \bibinfo{title}{State reconstruction of a multimode twin beam using
  photodetection}.
\newblock \emph{\bibinfo{journal}{Phys. Rev. A}} \textbf{\bibinfo{volume}{87}},
  \bibinfo{pages}{022108} (\bibinfo{year}{2013}).

\bibitem{Arkhipov2016e}
\bibinfo{author}{Arkhipov, I.~I.}, \bibinfo{author}{{Pe\v{r}ina~Jr.}, J.},
  \bibinfo{author}{Haderka, O.} \& \bibinfo{author}{Mich\'{a}lek, V.}
\newblock \bibinfo{title}{Experimental detection of nonclassicality of
  single-mode fields via intensity moments}.
\newblock \emph{\bibinfo{journal}{Opt. Express}} \textbf{\bibinfo{volume}{24}},
  \bibinfo{pages}{29496--29505} (\bibinfo{year}{2016}).

\bibitem{PerinaJr2017}
\bibinfo{author}{{Pe\v{r}ina~Jr.}, J.}, \bibinfo{author}{{Mich\' alek}, V.} \&
  \bibinfo{author}{Haderka, O.}
\newblock \bibinfo{title}{Higher-order sub-{P}oissonian-like nonclassical
  fields: {T}heoretical and experimental comparison}.
\newblock \emph{\bibinfo{journal}{Phys. Rev. A}} \textbf{\bibinfo{volume}{96}},
  \bibinfo{pages}{033852} (\bibinfo{year}{2017}).

\bibitem{PerinaJr2017a}
\bibinfo{author}{{Pe\v{r}ina~Jr.}, J.}, \bibinfo{author}{Arkhipov, I.~I.},
  \bibinfo{author}{{Mich\' alek}, V.} \& \bibinfo{author}{Haderka, O.}
\newblock \bibinfo{title}{Nonclassicality and entanglement criteria for
  bipartite optical fields characterized by quadratic detectors}.
\newblock \emph{\bibinfo{journal}{Phys. Rev. A}} \textbf{\bibinfo{volume}{96}},
  \bibinfo{pages}{043845} (\bibinfo{year}{2017}).

\bibitem{Achilles2003}
\bibinfo{author}{Achilles, D.}, \bibinfo{author}{Silberhorn, C.},
  \bibinfo{author}{\'{S}liwa, C.}, \bibinfo{author}{Banaszek, K.} \&
  \bibinfo{author}{Walmsley, I.~A.}
\newblock \bibinfo{title}{Fiber-assisted detection with photon number
  resolution}.
\newblock \emph{\bibinfo{journal}{Opt. Lett.}} \textbf{\bibinfo{volume}{28}},
  \bibinfo{pages}{2387--2389} (\bibinfo{year}{2003}).

\bibitem{Fitch2003}
\bibinfo{author}{Fitch, M.~J.}, \bibinfo{author}{Jacobs, B.~C.},
  \bibinfo{author}{Pittman, T.~B.} \& \bibinfo{author}{Franson, J.~D.}
\newblock \bibinfo{title}{Photon-number resolution using time-multiplexed
  single-photon detectors}.
\newblock \emph{\bibinfo{journal}{Phys. Rev. A}} \textbf{\bibinfo{volume}{68}},
  \bibinfo{pages}{043814} (\bibinfo{year}{2003}).

\bibitem{Haderka2004}
\bibinfo{author}{Haderka, O.}, \bibinfo{author}{Hamar, M.} \&
  \bibinfo{author}{{Pe\v{r}ina~Jr.}, J.}
\newblock \bibinfo{title}{Experimental multi-photon-resolving detector using a
  single avalanche photodiode}.
\newblock \emph{\bibinfo{journal}{Eur. Phys. J. D}}
  \textbf{\bibinfo{volume}{28}}, \bibinfo{pages}{149---154}
  (\bibinfo{year}{2004}).

\bibitem{Avenhaus2010}
\bibinfo{author}{Avenhaus, M.}, \bibinfo{author}{Laiho, K.},
  \bibinfo{author}{Chekhova, M.~V.} \& \bibinfo{author}{Silberhorn, C.}
\newblock \bibinfo{title}{Accessing higher order correlations in quantum
  optical states by time multiplexing}.
\newblock \emph{\bibinfo{journal}{Phys. Rev. Lett.}}
  \textbf{\bibinfo{volume}{104}}, \bibinfo{pages}{063602}
  (\bibinfo{year}{2010}).

\bibitem{Sperling2015}
\bibinfo{author}{Sperling, J.} \emph{et~al.}
\newblock \bibinfo{title}{Uncovering quantum correlations with time-multiplexed
  click detection}.
\newblock \emph{\bibinfo{journal}{Phys. Rev. Lett.}}
  \textbf{\bibinfo{volume}{115}}, \bibinfo{pages}{023601}
  (\bibinfo{year}{2015}).

\bibitem{Mosset2004}
\bibinfo{author}{Mosset, A.}, \bibinfo{author}{Devaux, F.},
  \bibinfo{author}{Fanjoux, G.} \& \bibinfo{author}{Lantz, E.}
\newblock \bibinfo{title}{Direct experimental characterization of the
  {Bose-Einstein} distribution of spatial fluctuations of spontaneous
  parametric down-conversion}.
\newblock \emph{\bibinfo{journal}{Eur. Phys. J. D - Atomic, Molecular, Opt.
  Plasma Phys.}} \textbf{\bibinfo{volume}{28}}, \bibinfo{pages}{447---451}
  (\bibinfo{year}{2004}).

\bibitem{Blanchet2008}
\bibinfo{author}{Blanchet, J.-L.}, \bibinfo{author}{Devaux, F.},
  \bibinfo{author}{Furfaro, L.} \& \bibinfo{author}{Lantz, E.}
\newblock \bibinfo{title}{Measurement of sub-shot-noise correlations of spatial
  fluctuations in the photon-counting regime}.
\newblock \emph{\bibinfo{journal}{Phys. Rev. Lett.}}
  \textbf{\bibinfo{volume}{101}}, \bibinfo{pages}{233604}
  (\bibinfo{year}{2008}).

\bibitem{PerinaJr2012}
\bibinfo{author}{{Pe\v{r}ina~Jr.}, J.}, \bibinfo{author}{Hamar, M.},
  \bibinfo{author}{Mich\'{a}lek, V.} \& \bibinfo{author}{Haderka, O.}
\newblock \bibinfo{title}{Photon-number distributions of twin beams generated
  in spontaneous parametric down-conversion and measured by an intensified
  {CCD} camera}.
\newblock \emph{\bibinfo{journal}{Phys. Rev. A}} \textbf{\bibinfo{volume}{85}},
  \bibinfo{pages}{023816} (\bibinfo{year}{2012}).

\bibitem{PerinaJr2017b}
\bibinfo{author}{{Pe\v{r}ina~Jr.}, J.}, \bibinfo{author}{{Mich\' alek}, V.} \&
  \bibinfo{author}{Haderka, O.}
\newblock \bibinfo{title}{Noise reduction in photon counting by exploiting
  spatial correlations}.
\newblock \emph{\bibinfo{journal}{Phys. Rev. Appl.}}
  \textbf{\bibinfo{volume}{8}}, \bibinfo{pages}{044018} (\bibinfo{year}{2017}).

\bibitem{Machulka2014}
\bibinfo{author}{Machulka, R.} \emph{et~al.}
\newblock \bibinfo{title}{Spatial properties of twin-beam correlations at low-
  to high-intensity transition}.
\newblock \emph{\bibinfo{journal}{Opt. Express}} \textbf{\bibinfo{volume}{22}},
  \bibinfo{pages}{13374---13379} (\bibinfo{year}{2014}).

\bibitem{PerinaJr2014a}
\bibinfo{author}{{Pe\v{r}ina~Jr.}, J.}, \bibinfo{author}{Haderka, O.},
  \bibinfo{author}{Allevi, A.} \& \bibinfo{author}{Bondani, M.}
\newblock \bibinfo{title}{Absolute calibration of photon-number-resolving
  detectors with an analog output using twin beams}.
\newblock \emph{\bibinfo{journal}{Appl. Phys. Lett.}}
  \textbf{\bibinfo{volume}{104}}, \bibinfo{pages}{041113}
  (\bibinfo{year}{2014}).

\bibitem{Ramilli2010}
\bibinfo{author}{Ramilli, M.} \emph{et~al.}
\newblock \bibinfo{title}{Photon-number statistics with silicon
  photomultipliers}.
\newblock \emph{\bibinfo{journal}{J. Opt. Soc. Am. B}}
  \textbf{\bibinfo{volume}{27}}, \bibinfo{pages}{852---862}
  (\bibinfo{year}{2010}).

\bibitem{Paris97}
\bibinfo{author}{Paris, M.~G.~A.}
\newblock \bibinfo{title}{Joint generation of identical squeezed states}.
\newblock \emph{\bibinfo{journal}{Phys. Lett. A}}
  \textbf{\bibinfo{volume}{225}}, \bibinfo{pages}{28} (\bibinfo{year}{1997}).

\bibitem{Arkhipov2016c}
\bibinfo{author}{Arkhipov, I.~I.}, \bibinfo{author}{{Pe\v{r}ina~Jr.}, J.},
  \bibinfo{author}{Pe\v{r}ina, J.} \& \bibinfo{author}{Miranowicz, A.}
\newblock \bibinfo{title}{Interplay of nonclassicality and entanglement of
  two-mode {G}aussian fields generated in optical parametric processes}.
\newblock \emph{\bibinfo{journal}{Phys. Rev. A}} \textbf{\bibinfo{volume}{94}},
  \bibinfo{pages}{013807} (\bibinfo{year}{2016}).

\bibitem{Arkhipov2016b}
\bibinfo{author}{Arkhipov, I.~I.}, \bibinfo{author}{{Pe\v{r}ina~Jr.}, J.},
  \bibinfo{author}{Svozil\'ik, J.} \& \bibinfo{author}{Miranowicz, A.}
\newblock \bibinfo{title}{Nonclassicality invariant of general two-mode
  {G}aussian states}.
\newblock \emph{\bibinfo{journal}{Sci. Rep.}} \textbf{\bibinfo{volume}{6}},
  \bibinfo{pages}{26523} (\bibinfo{year}{2016}).

\bibitem{Glauber1963}
\bibinfo{author}{Glauber, R.~J.}
\newblock \bibinfo{title}{Coherent and incoherent states of the radiation
  field}.
\newblock \emph{\bibinfo{journal}{Phys. Rev.}} \textbf{\bibinfo{volume}{131}},
  \bibinfo{pages}{2766---2788} (\bibinfo{year}{1963}).

\bibitem{Sudarshan1963}
\bibinfo{author}{Sudarshan, E. C.~G.}
\newblock \bibinfo{title}{Equivalence of semiclassical and quantum mechanical
  descriptions of statistical light beams}.
\newblock \emph{\bibinfo{journal}{Phys. Rev. Lett.}}
  \textbf{\bibinfo{volume}{10}}, \bibinfo{pages}{277} (\bibinfo{year}{1963}).

\bibitem{Shchukin2005}
\bibinfo{author}{Shchukin, E.}, \bibinfo{author}{Richter, T.} \&
  \bibinfo{author}{Vogel, W.}
\newblock \bibinfo{title}{Nonclassicality criteria in terms of moments}.
\newblock \emph{\bibinfo{journal}{Phys. Rev. A}} \textbf{\bibinfo{volume}{71}},
  \bibinfo{pages}{011802(R)} (\bibinfo{year}{2005}).

\bibitem{Miranowicz2010}
\bibinfo{author}{Miranowicz, A.}, \bibinfo{author}{Bartkowiak, M.},
  \bibinfo{author}{Wang, X.}, \bibinfo{author}{Liu, X.-Y.} \&
  \bibinfo{author}{Nori, F.}
\newblock \bibinfo{title}{Testing nonclassicality in multimode fields: A
  unified derivation of classical inequalities}.
\newblock \emph{\bibinfo{journal}{Phys. Rev. A}} \textbf{\bibinfo{volume}{82}},
  \bibinfo{pages}{013824} (\bibinfo{year}{2010}).

\bibitem{Bartkowiak2011b}
\bibinfo{author}{Bartkowiak, M.} \emph{et~al.}
\newblock \bibinfo{title}{Sudden vanishing and reappearance of nonclassical
  effects: General occurrence of finite-time decays and periodic vanishings of
  nonclassicality and entanglement witnesses}.
\newblock \emph{\bibinfo{journal}{Phys. Rev. A}} \textbf{\bibinfo{volume}{83}},
  \bibinfo{pages}{053814} (\bibinfo{year}{2011}).

\bibitem{Lee1990a}
\bibinfo{author}{Lee, C.~T.}
\newblock \bibinfo{title}{Higher-order criteria for nonclassical effects in
  photon statistics}.
\newblock \emph{\bibinfo{journal}{Phys. Rev. A}} \textbf{\bibinfo{volume}{41}},
  \bibinfo{pages}{1721---1723} (\bibinfo{year}{1990}).

\bibitem{PerinaKrepelka11}
\bibinfo{author}{{Pe\v{r}ina}, J.} \& \bibinfo{author}{{K\v{r}epelka}, J.}
\newblock \bibinfo{title}{Joint probability distribution and entanglement in
  optical parametric processes}.
\newblock \emph{\bibinfo{journal}{Opt. Commun.}}
  \textbf{\bibinfo{volume}{284}}, \bibinfo{pages}{4941} (\bibinfo{year}{2011}).

\bibitem{PerinaKrepelka05}
\bibinfo{author}{Pe\v{r}ina, J.} \& \bibinfo{author}{K\v{r}epelka, J.}
\newblock \bibinfo{title}{Multimode description of spontaneous parametric
  down-conversion}.
\newblock \emph{\bibinfo{journal}{J. Opt. B: Quantum Semiclass. Opt.}}
  \textbf{\bibinfo{volume}{7}}, \bibinfo{pages}{246} (\bibinfo{year}{2005}).

\bibitem{Arkhipov15}
\bibinfo{author}{Arkhipov, I.~I.}, \bibinfo{author}{{Pe\v{r}ina~Jr.}, J.},
  \bibinfo{author}{Pe\v{r}ina, J.} \& \bibinfo{author}{Miranowicz, A.}
\newblock \bibinfo{title}{Comparative study of nonclassicality, entanglement,
  and dimensionality of multimode noisy twin beams}.
\newblock \emph{\bibinfo{journal}{Phys. Rev. A}} \textbf{\bibinfo{volume}{91}},
  \bibinfo{pages}{033837} (\bibinfo{year}{2015}).

\bibitem{Plenio05}
\bibinfo{author}{Plenio, M.~B.}
\newblock \bibinfo{title}{Logarithmic negativity: A full entanglement monotone
  that is not convex}.
\newblock \emph{\bibinfo{journal}{Phys. Rev. Lett.}}
  \textbf{\bibinfo{volume}{95}}, \bibinfo{pages}{090503}
  (\bibinfo{year}{2005}).

\bibitem{Harder2016}
\bibinfo{author}{Harder, G.} \emph{et~al.}
\newblock \bibinfo{title}{Single-mode parametric-down-conversion states with 50
  photons as a source for mesoscopic quantum optics}.
\newblock \emph{\bibinfo{journal}{Phys. Rev. Lett.}}
  \textbf{\bibinfo{volume}{116}}, \bibinfo{pages}{143601}
  (\bibinfo{year}{2016}).

\end{thebibliography}

\noindent {\bf Acknowledgments} The authors were supported by M\v{S}MT \v{C}R
(Project No.~LO1305) and GA \v{C}R (J.P.Jr: Project No.~15-08971S,
I.A.: Project No.~17-23005Y).

\noindent {\bf Author contributions statement} I.A. and J.P. developed the theory and wrote the manuscript. I.A. prepared  figures.

\noindent {\bf Additional information}

\textbf{Accession codes};

\textbf{Competing financial interests} The authors have no
competing financial interests.

\end{document}